\documentclass[sigconf]{acmart}

\usepackage{booktabs}
\usepackage{arydshln}
\usepackage{multirow}
\usepackage{float} 
\usepackage{xcolor}
\usepackage{colortbl}

\AtBeginDocument{%
  }

\copyrightyear{2025}
\acmYear{2025}
\setcopyright{rightsretained}
\acmConference[RecSys '25]{Proceedings of the Nineteenth ACM Conference on Recommender Systems}{September 22--26, 2025}{Prague, Czech Republic}
\acmBooktitle{Proceedings of the Nineteenth ACM Conference on Recommender Systems (RecSys '25), September 22--26, 2025, Prague, Czech Republic}\acmDOI{10.1145/3705328.3759305}
\acmISBN{979-8-4007-1364-4/2025/09}
\begin{document}

\title{Evaluating Podcast Recommendations with Profile-Aware LLM-as-a-Judge}

\author{Francesco Fabbri\texorpdfstring{\thanks{Corresponding author: francescof@spotify.com}}{}}
\affiliation{%
  \institution{Spotify}
  \country{Spain}
}

\author{Gustavo Penha}
\affiliation{%
  \institution{Spotify}
  \country{Netherlands}
}

\author{Edoardo D'Amico}
\affiliation{%
  \institution{Spotify}
  \country{Spain}
}

\author{Alice Wang}
\affiliation{%
  \institution{Spotify}
  \country{United States}
}

\author{Marco De Nadai}
\affiliation{%
  \institution{Spotify}
  \country{Denmark}
}

\author{Jackie Doremus}
\affiliation{%
  \institution{Spotify}
  \country{United States}
}

\author{Paul Gigioli}
\affiliation{%
  \institution{Spotify}
  \country{United States}
}

\author{Andreas Damianou}
\affiliation{%
  \institution{Spotify}
  \country{United Kingdom}
}

\author{Oskar Stål}
\affiliation{%
  \institution{Spotify}
  \country{Sweden}
}

\author{Mounia Lalmas}
\affiliation{%
  \institution{Spotify}
  \country{United Kingdom}
}

\renewcommand{\shortauthors}{Fabbri et al.}

\begin{abstract}

Evaluating personalized recommendations remains a central challenge, especially in long-form audio domains like podcasts, where traditional offline metrics suffer from exposure bias and online methods such as A/B testing are costly and operationally constrained. In this paper, we propose a novel framework that leverages Large Language Models (LLMs) as offline judges to assess the quality of podcast recommendations in a scalable and interpretable manner. Our two-stage profile-aware approach first constructs natural-language user profiles distilled from 90 days of listening history. These profiles summarize both topical interests and behavioral patterns, serving as compact, interpretable representations of user preferences. Rather than prompting the LLM with raw data, we use these profiles to provide high-level, semantically rich context—enabling the LLM to reason more effectively about alignment between a user’s interests and recommended episodes. This reduces input complexity and improves interpretability. The LLM is then prompted to deliver fine-grained pointwise and pairwise judgments based on the profile-episode match. In a controlled study with 47 participants, our profile-aware judge matched human judgments with high fidelity and outperformed or matched a variant using raw listening histories. The framework enables efficient, profile-aware evaluation for iterative testing and model selection in recommender systems.

\end{abstract}

\ccsdesc[500]{Information systems~Language models}
\ccsdesc[500]{Information systems~Personalization}

\maketitle

\section{Introduction}

Evaluating personalized recommender systems remains a fundamental challenge, largely due to the limitations of offline evaluations methods and metrics~\cite{thomas2024llmsearchprefs}. Standard metrics like hit rate and recall are based on historical interaction data, which introduces exposure bias: models are evaluated only on items users have previously seen, not the full space of potential recommendations. This makes it difficult to accurately assess a model’s true effectiveness.

These shortcomings are especially pronounced in cold-start scenarios, such as the introduction of new features (e.g., a new podcast shelf), where no historical interaction data exists. In such cases, offline metrics fail, and practitioners must rely on qualitative assessments to estimate alignment with the intended user experience before launch. At the other extreme, A/B testing and user studies, while grounded in real behavior, are costly, slow, and operationally constrained, limiting the number of models that can be practically tested. As a result, practitioners face a dilemma: fast but limited offline evaluation, or rigorous but slow experimentation. This reveals a critical gap: \emph{the lack of a scalable, reliable middle ground for pre-deployment model selection}.

Traditional evaluation methods, whether quantitative or qualitative, also fall short in capturing true user satisfaction or explaining why a recommendation is relevant. Crucially, they fail to determine whether a recommendation meaningfully reflects a user's underlying preferences. This challenge is especially acute in the podcast domain, where the cost of a poor recommendation is high~\cite{jones2021podcast}; unlike short-form content, podcasts require considerable attention. Implicit feedback, such as stopping after ten minutes, can signal strong disinterest, mild curiosity, or simple distraction, making interpretation highly ambiguous.

Unlike search, where evaluation checks whether retrieved results satisfy an explicit user query, recommendation must infer intent entirely from behavioral traces. In search, the query serves as a content hypothesis, a direct expression of the user's information need. In verticals like “music from the 80s,” the scope is often predefined by the domain or interface. But in personalized recommendation, especially for long-form content, no such explicit formulation of user intent exists. This challenge is particularly acute in podcast recommendation, where user preferences span multiple dimensions—including topic, tone, format, and host style—and are difficult to infer from sparse interaction data. The core evaluation task, therefore, becomes one of constructing a content hypothesis: an interpretable approximation of what the user prefers, inferred from past listening behavior.

We propose that this missing hypothesis can be explicitly constructed in the form of a natural-language user profile: a structured summary of topical interests, stylistic preferences, and behavioral patterns distilled from listening history. These profiles provide high-level, interpretable context that allows Large Language Models (LLMs) to reason more effectively about whether a recommendation aligns with inferred user intent.

LLMs offer a promising path forward for scalable, human-aligned evaluation~\cite{gu2025surveyllmasajudge}. Models like GPT-4~\cite{vaswani2017attention, brown2020language} show high agreement with human judgments across diverse tasks~\cite{xu2025contextualjudgebench}, and the “LLM-as-a-Judge” paradigm is emerging as a general evaluation strategy~\cite{ye2025justice}.
LLMs can assess relevance in relation to user preferences~\cite{wang2024llms, wang2024large}. However, prior work often feeds raw interaction data to the LLM or assumes structured ground-truth signals, limiting interpretability.

Recent work on personalized judges~\cite{dong-etal-2024-llm} highlights the limitations of generic LLM-based evaluation when user context is under-specified. This underscores the need for profile-aware prompting strategies that encode nuanced, personalized context. We argue that structured, profile-based representations enable more faithful alignment evaluation, and unlock the full potential of LLMs as offline judges for personalized systems, especially for pre-deployment settings, where traditional online experimentation is too costly, slow, or operationally infeasible. 

\paragraph{Our Approach} To address this challenge, we introduce a profile-aware LLM-as-a-Judge framework (\emph{Judge} throughout the paper) for evaluating personalized podcast episode recommendations. Central to our framework is a natural-language profile automatically distilled from each user’s listening history, which serves as an explicit content hypothesis representing the user’s inferred preferences. The LLM is prompted with this profile and candidate episode metadata to reason about alignment along multiple dimensions, such as topic, tone, and format.
The framework supports two complementary evaluation modes:
\begin{enumerate}
    \item Pointwise evaluation: the \emph{Judge} assesses whether an individual episode aligns with the user’s inferred preferences.
    \item Pairwise evaluation: in a setup analogous to A/B testing, the \emph{Judge} compares two ranked episode lists, each from a different model, and select the one better aligned with the profile.
\end{enumerate}

Together, these evaluation modes offer a scalable, interpretable mechanism for judging recommendation quality, bridging the gap between coarse offline metrics and more subjective, human-aligned assessments of user satisfaction.

\section{Related Work}
\label{sec:related}

Recent work has formalized the use of LLMs as evaluators of system outputs, a methodology widely referred to as LLM-as-a-Judge. Originally developed for dialogue evaluation and instruction following~\citep{zheng2023mtbench, fu-etal-2024-gptscore}, this paradigm has since expanded across domains,  leading to structured evaluation toolkits and taxonomies~\citep{gu2025surveyllmasajudge, xu2025contextualjudgebench, huang2024foundation, lin2025can}.

For instance, \citet{zheng2023mtbench} proposed large-scale preference datasets that surface biases related to response position and verbosity. \citet{fu-etal-2024-gptscore} demonstrated that instruction-tuned LLMs can act as flexible, robust scorers of generation quality. More recently, \citet{gu2025surveyllmasajudge} surveyed key tasks, prompting strategies, and open challenges, while~\citet{xu2025contextualjudgebench} emphasized the importance of supplying relevant user context for reliable judgment.

Concerns about bias and misalignment have also been raised. \citet{ye2025justice} cataloged systematic biases in LLM judgments, and~\citet{sahoo2025quantitativellmjudges} propose post-hoc regression  calibration techniques. \citet{thakur2025judging} highlighted gaps in alignment and prompt sensitivity across judge models. While our work focuses on profile-based alignment, we do not explicitly address these issues. Investigating bias mitigation and prompt robustness is an important direction for future work.

In information retrieval, \citet{thomas2024llmsearchprefs} showed that GPT-4 can predict document relevance with near-human accuracy. However, applying LLMs to recommendation introduces additional challenges: user preferences must be inferred from behavior over time, and recommendations lack an explicit query to ground evaluation. Our framework addresses this by constructing natural-language profiles that act as explicit content hypotheses: structured representations of inferred user intent, enabling LLMs to evaluate alignment in personalized, dynamic settings.

Related work by \citet{dong-etal-2024-llm} found that persona-conditioned prompting improves evaluation in dialogue tasks. Our approach differs in two ways: $(i)$ we apply LLM-based judgment to ranking in personalized recommendation rather than conversation quality, and $(ii)$ our user profiles are automatically distilled from behavioral traces, not manually crafted.

Finally, \citet{jones2021podcast} highlighted a lack of scalable offline evaluation methods for podcast recommendations. Our work directly addresses this gap by introducing a profile-aware LLM-as-a-Judge framework, along with open tools to support evaluation in long-form, preference-driven media domains.

\section{LLM-as-a-Judge for Offline Testing}

Evaluating podcast recommendations poses unique challenges due to the nuanced, multi-dimensional nature of user satisfaction. Traditional methods typically rely on observable behavior, but in long-form audio contexts, such signals are difficult to interpret. Implicit feedback is sparse and often ambiguous, and standard metrics fail to capture the richness of listener preferences. Irregular consumption patterns and the high time cost of engagement further limit the reliability of behavioral signals, creating significant evaluation gaps. Although prior work has called for richer evaluation frameworks, few offer scalable solutions for detecting recommendation misalignment~\cite{jones2021podcast}.

To address this, we introduce a \emph{profile-aware} evaluation framework that leverages LLMs as interpretable, domain-adaptive offline judges. Rather than relying on item-level engagement signals (e.g., clicks or listens), our approach uses structured natural-language profiles distilled from listening history to assess how well a recommended episode aligns with a user’s topical interests and behavioral patterns. This bridges the gap between nuanced relevance criteria and the scalability needs of offline evaluation, offering a practical alternative to coarse numerical proxies.

\begin{figure}[t]
    \centering
    \includegraphics[width=1\linewidth]{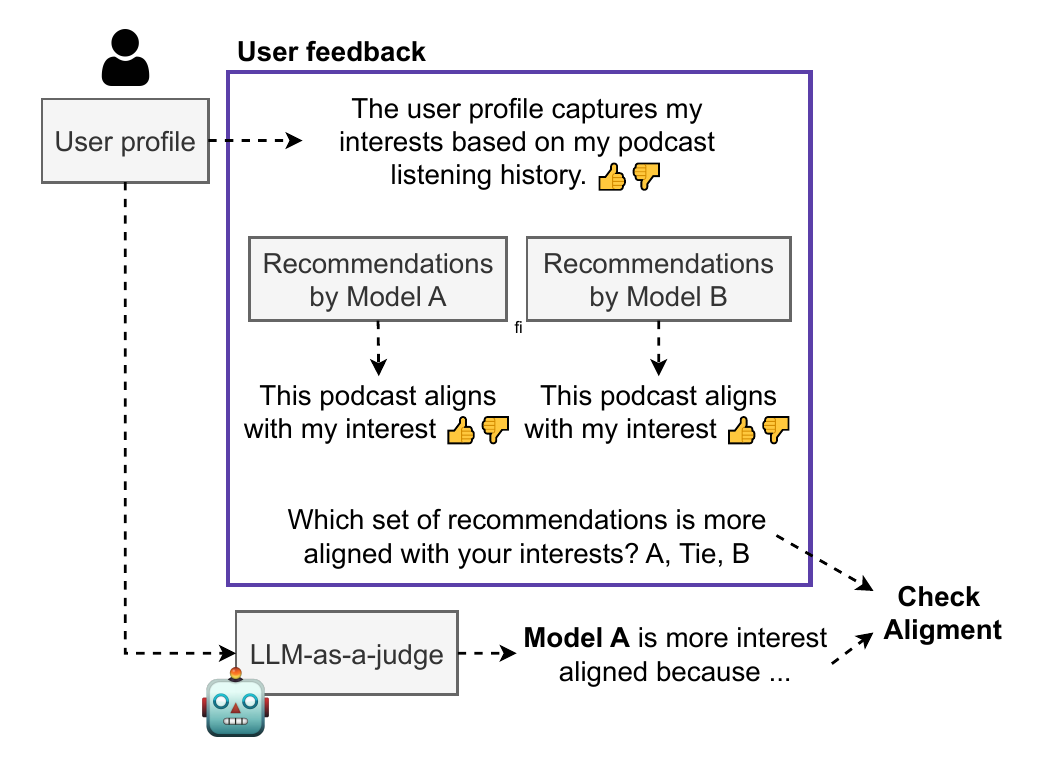}
    \caption{LLM-as-a-Judge evaluation pipeline. The system takes as input a user profile synthesized from listening history and two sets of recommended episodes, and outputs rationales and binary judgments for episode-level fit and model-level comparison.}
    \label{fig:enter-label}
      \vspace*{-0.1in}
\end{figure}

\begin{figure}[t]
    \centering
    \includegraphics[width=1\linewidth]{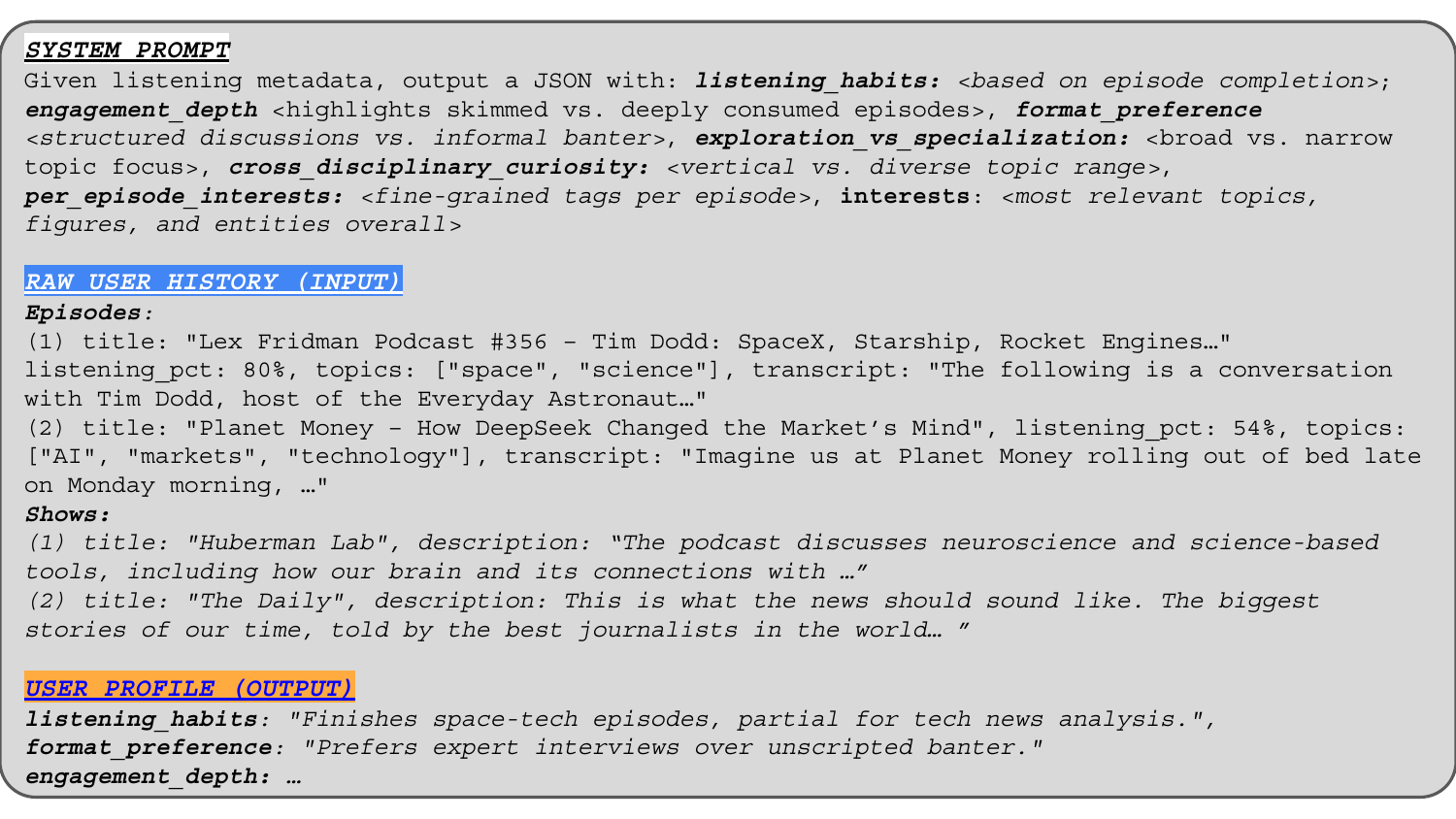}
    \caption{Profile generation prompt. The LLM receives structured listening metadata and is prompted to produce a natural-language user profile with interpretable dimensions (e.g., listening habits, format preference). This profile is later used as input for evaluating recommendation relevance.}
    \label{fig:user-prompt}
\end{figure}

The framework operates in two key stages: user profiling and episode assessment. In the first stage, we generate a structured profile for each user based on their most recent three months of listening activity. This profile is derived from podcast metadata (including titles, descriptions, transcripts, and topical tags) associated with episodes and shows the user has engaged with most. The profile captures two main dimensions:
\begin{itemize}
    \item {\bf Content preferences:} topical and named-entity focus, cross-domain curiosity, and tendencies toward exploration or specialization
    \item {\bf Listening patterns:} habits, engagement depth, and format preferences.
\end{itemize}
These six attributes form a comprehensive user representation, which is then used for evaluating alignment with candidate episodes (Fig.~\ref{fig:user-prompt}).

In the second stage the \emph{Judge}, an off-the-shelf LLM queried in zero-shot mode (i.e., no fine-tuning or calibration), is prompted with both the user profile and the metadata of a recommended episode. Using a Chain-of-Thought reasoning style, the \emph{Judge} produces a rationale and a binary judgment indicating whether the episode is a good fit; this constitutes the pointwise evaluation~\cite{wei2022chain}. While we also tested a multiclass version (including neutral feedback), it yielded no substantial improvement and is omitted for brevity.

For model-level evaluation, the \emph{Judge} performs pairwise comparisons between two ranked lists of episodes, each generated by a different recommendation model, and selects the list that better matches the user profile. This setup is designed to compare models with different architectures but similar optimization goals. For each comparison, the \emph{Judge} provides: (1) dimension-wise qualitative rationale outlining the strengths and weaknesses of each list; and (2) a final verdict, either preferring one model or indicating a tie when neither shows clear superiority. To mitigate position bias, the identity tags of Model A and Model B are randomly shuffled before each evaluation, ensuring an unbiased and reliable comparison.

\section{Experiments}

\paragraph{Setup} To evaluate the validity of our framework, referred to throughout this section as \textbf{LaaJ} (\emph{LLM-as-a-Judge}), we conduct a controlled experiment with real users to assess whether it can serve as a reliable offline judge on recommendation quality. The experiment involved two anonymized models (Model A and Model B), 47 participants, and a two-stage evaluation comparing LLM-generated judgments with human feedback per user. Each participant first receives a personalized profile, automatically generated from their podcast listening history. 

Then, two sets of episode recommendations are generated, one for each model, and displayed side-by-side. Each set includes 3 episodes per model, with each episode shown alongside its  show name, description, cover image, and a playable audio segment.
Through the survey interface, users can provided structured feedback on: (1) the accuracy of their profile; (2) how well each episode aligns with their interests; (3) which model better matches their preferences overall.

Participants rated each item using a 5-point Likert scale: \emph{Strongly Disagree, Disagree, Neutral, Agree, and Strongly Agree}. To ensure unbiased feedback, all participants were blinded to the identity of the models. The LLM Judge was prompted using static templates, with no fine-tuning or post-hoc calibration applied~\cite{zhang2025cold}.

The compared models differ in architecture: (1) {\bf Model A}, which was primarily content-based, with less sensitivity to consumption patterns, and (2) {\bf Model B}, which relied heavily on collaborative filtering signals, with limited content-based integration. For all experiments we used \emph{GPT-4.1}~\cite{achiam2023gpt} for both the user profile generator and the judging model. We chose it  for its reported strong alignment with human preferences, consistent performance across evaluation tasks, and better correlation with human judgments than other LLMs~\cite{zheng2023mtbench,liu-etal-2023-g}.

\paragraph{LLM Agreement \& Judgment Behavior} We present a comparison between the output of the \emph{Judge} and the human-annotated data. From 47 users included in the study, we collect in total 277 pointwise human evaluations and 47 model-level comparisons (one per user). The dataset covers 227 unique recommended episodes, with an average of 5.89 episode annotations per user.

In our evaluation we test three different judges, including two \emph{LaaJ} variants, and a  non-LLM one: 
\begin{itemize}
    \item \textbf{LaaJ-Profile (our profile-aware judge)}: uses a structured, natural-language summary of each user’s listening history, distilled from their top shows and episodes. Profiles serve as an interpretable content hypothesis, capturing topical preferences, stylistic traits, and behavioral patterns, to guide the LLM’s reasoning about alignment, without requiring access to raw interaction data.
    \item \textbf{LaaJ-History}: a variant that provides the LLM with the full set of shows and episodes from the user’s listening history, rather than a distilled profile. This approach tests whether reasoning directly over raw behavioral traces leads to better alignment judgments, and serves as a baseline for evaluating the benefits of compressing user preferences into a structured, interpretable profile.
    \item \textbf{sBERT-Sim}: a non-LLM baseline that computes cosine similarity between Sentence-BERT embeddings of the user profile and episode metadata. Episodes are marked aligned if similarity exceeds a fixed threshold (0.5), and model-level alignment is determined by aggregating episode-level scores. This serves as a simple, interpretable proxy for content-level user-item relevance.
    
\end{itemize}
\begin{table}[t]
  \centering\small
  \begin{tabular}{lccc}
    \toprule
    \textbf{Model} & \textbf{ROC-AUC} & \textbf{MSA (W/T/L)} & \textbf{RSM} \\
    \midrule
    LaaJ-Profile  & 0.6442 & 0.6596 (30/1/16) & 0.6667 \\
    LaaJ-History  & 0.6476 & 0.6170 (28/1/18) & 0.6667 \\
    sBERT-Sim      & 0.4871 & 0.5106 (21/3/23)                        & 0.5000 \\
    \bottomrule
  \end{tabular}
  \caption{Performance on both episode and model evaluations of different judges on the human-labeled  dataset.  “W/T/L” counts wins, ties, and losses against the human label.}
   \label{table1}
   \vspace*{-0.1in}
\end{table}

Table~\ref{table1} presents results from both episode-level and model-level evaluations. The pointwise evaluation is conducted on 277 annotated episodes. We report the following metrics: ROC-AUC measures the accuracy of the judges on user-episode predictions; Model Selection Agreement (MSA) is the fraction of cases where the judge's model preference matches annotators’ choices; Outcome Distribution is the number of Wins, Ties, and Losses for the judge compared to ground-truth human annotations; and Recall of Strong Misalignment (RSM) is the proportion of strongly misaligned recommendations flagged by the judges also identified by annotators as clearly misaligned with user preferences.

As shown Table~\ref{table1}, \emph{LaaJ-Profile} achieves comparable ROC-AUC to \emph{LaaJ-History}, despite relying solely on a natural-language profile rather than the user history. This demonstrates that a concise, interpretable representation of user preferences can serve as an effective content hypothesis that captures the essence of user intent. In model-level comparisons, the profile-based variant outperforms the history-based, underscoring the value of summarizing multi-faceted user interests for reliable comparative judgments between recommendation models. Additionally, both LLM-based judges correctly identify 66\% of strongly misaligned episodes (per the RSM metric), indicating sensitivity to recommendations that conflict with user preferences.

\begin{figure}[t]
    \centering
    \begin{tabular}{ll}
        \includegraphics[width=0.45\linewidth]{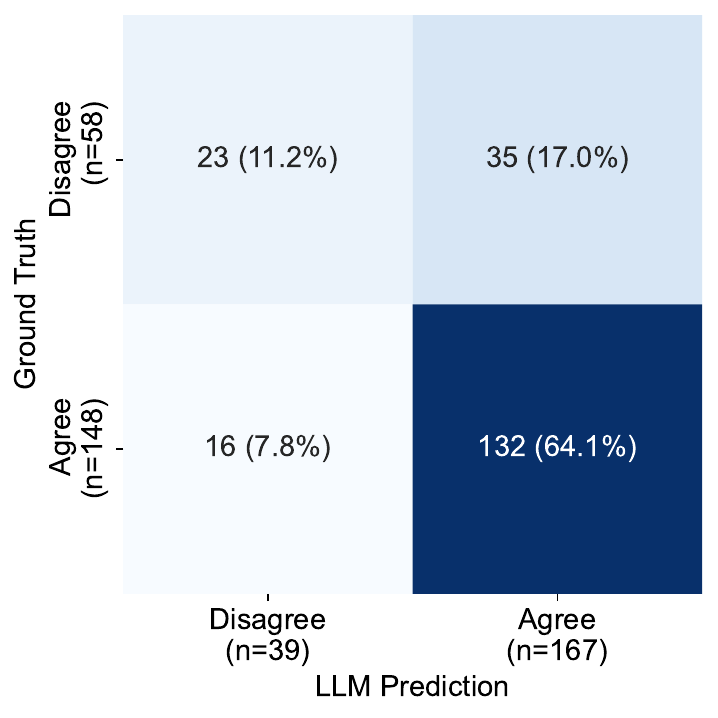} &
        \includegraphics[width=0.45\linewidth]{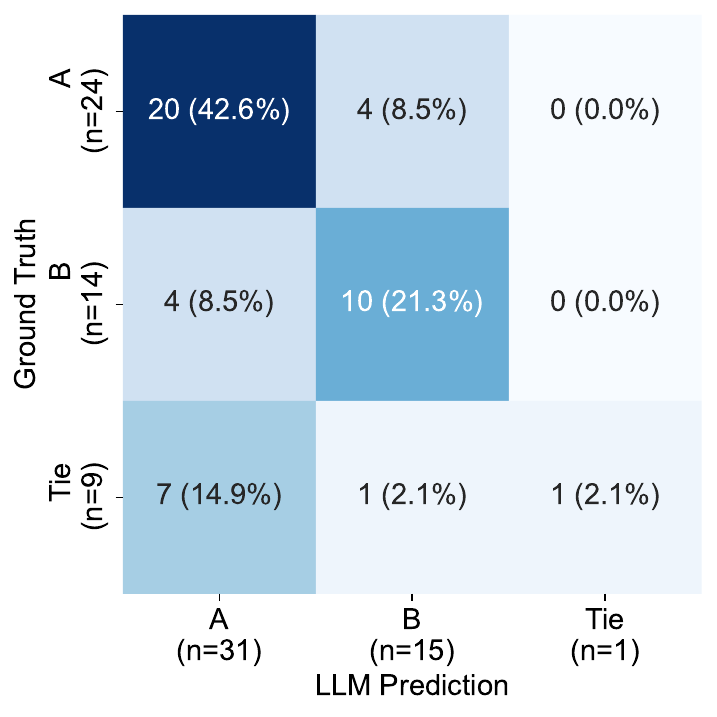}
\end{tabular}
\caption{Confusion matrices comparing the profile-aware \emph{LaaJ-Profile} with human annotations. Columns represent LLM decisions; rows show human relevance labels. Left: episode-level (pointwise) comparison. Right: model-level (pairwise) comparison.}
\label{heatmaps}
  \vspace*{-0.18in}
\end{figure}

Continuing our analysis of \emph{LaaJ-Profile}, we examine its confusion matrices against human annotations (Fig.~\ref{heatmaps}). In episode-level evaluation, the matrix (left) shows alignment in 75\% of the cases. However, 17\% of the episodes were judged as aligned by the LLM but not by users (representing false positives). This discrepancy reflects a known tendency of LLMs to produce positively skewed responses~\cite{zhujudgelm, gallegos2024bias}.

In the model-level (pairwise) evaluation, the confusion matrix (right) reveals strong agreement between the Judge and human annotators in preferring Model A over Model B, with 20 true positives out of 24 comparisons. However, the LLM tends to be more decisive: it registers only one tie, in contrast to the eight ties recorded by human annotators. This tendency may be addressed through more adaptive in-context learning strategies or by model fine-tuning~\cite{wang2025user}.

Qualitative feedback from human annotators revealed their judgments were influenced by factors beyond standard evaluation metrics, such as familiarity with the show, the identity of the host, stylistic tone, and the diversity of the recommendations. While some users preferred narrowly focused and familiar recommendation lists, others placed higher value on variety and novelty. These findings highlight the complex, multi-dimensional, and inherently subjective nature of podcast preferences in real-world settings.

\begin{figure}[t]
     \centering
     \begin{tabular}{ll}
     \includegraphics[width=0.47\linewidth]{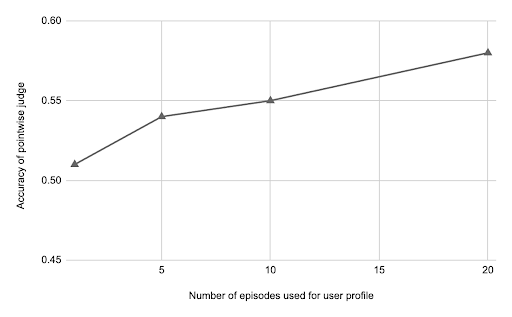}
    \includegraphics[width=0.47\linewidth]{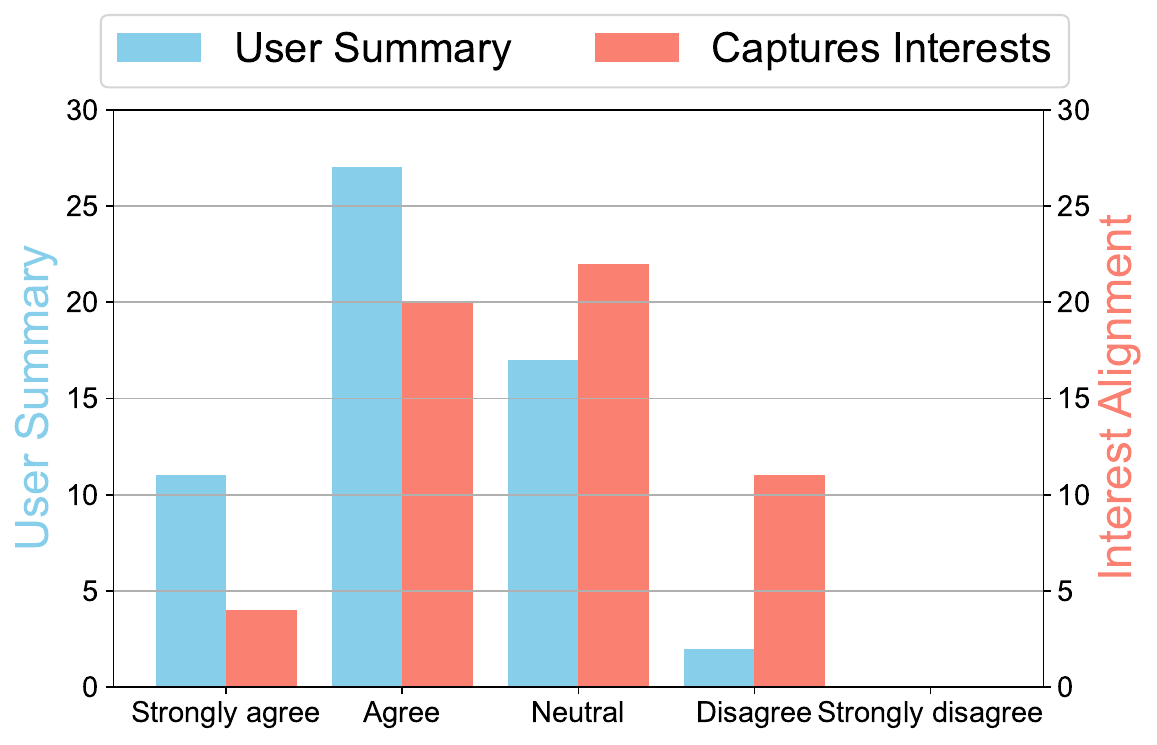}
     \end{tabular}
     \caption{Left: Impact of user profile length on \emph{LaaJ}-human alignment. On x-axis the number of episodes used to generate the profile; on y-axis the \emph{LaaJ}-human accuracy. Right: Human agreement on profile quality and interest alignment. Bar chart includes two frequency distributions: $(i)$ alignment with user preferences (blue);$(ii)$ alignment with users' interests (red).}
     \label{user-profile}
    \vspace*{-0.1in}   
 \end{figure}

\paragraph{Impact of User Profiles} Fig.~\ref{user-profile} shows that increasing the number of shows and episodes used to generate user profiles in \emph{LaaJ-Profile}  improves judgment accuracy, raising alignment with human preferences by +8\% from 0.51 with 5 episodes to 0.59 with 20 episodes. This emphasizes the critical role of context richness and profile coverage in enabling the LLM to make accurate evaluations.

Participants were asked to review their automatically generated profiles in \emph{LaaJ-Profile} and evaluate how well they reflected their listening preferences. As shown in Fig.~\ref{user-profile}, most agreed the profiles offered a reasonable high-level summary, but views were more divided on how accurately the profiles captured their deeper interests. Quantitative ratings indicated the profiles were broadly representative, yet qualitative feedback added nuance. While many users recognized that key aspects of their listening behavior were captured, some expressed concerns about the depth and specificity of representation. Some users pointed to missing personal elements such as favorite hosts, limited coverage of stylistic tone, and a narrow topical focus—often shaped by recent listening activity. 

These observations reflect the difficulty of inferring subjective preferences from short or sparse interaction histories, as well as the trade-off between recency and long-term interest modeling. Several participants noted that short-term data windows sometimes failed to reflect enduring tastes. These insights point to opportunities for enhancing profiles by incorporating long-term behavioral signals and more nuanced metadata.

\section{Conclusions \& Future Work}

This paper presents a scalable framework for using LLMs as offline judges to evaluate personalized podcast recommendations through the lens of user preference alignment. At the core of our approach are structured, natural-language profiles that act as explicit content hypotheses: interpretable summaries of likely user preferences distilled from listening history. Prompting LLMs with these profiles, rather than raw behavioral data, enables more accurate and interpretable alignment judgments at both episode and model levels. Our experiments show that this profile-aware evaluation matches or exceeds the performance of history-based alternatives. 

Looking ahead, we aim to improve profile fidelity by incorporating long-term behavior and explicit feedback~\cite{su2025regen}, and to explore adaptive prompting (e.g., few-shot or in-context learning) to enhance robustness and reduce decisiveness bias. We also plan to extend the approach across domains and user groups to assess its generalizability and impact at scale.

\bibliographystyle{ACM-Reference-Format}
\bibliography{main}


\begin{thebibliography}{24}


\ifx \showCODEN    \undefined \def \showCODEN     #1{\unskip}     \fi
\ifx \showISBNx    \undefined \def \showISBNx     #1{\unskip}     \fi
\ifx \showISBNxiii \undefined \def \showISBNxiii  #1{\unskip}     \fi
\ifx \showISSN     \undefined \def \showISSN      #1{\unskip}     \fi
\ifx \showLCCN     \undefined \def \showLCCN      #1{\unskip}     \fi
\ifx \shownote     \undefined \def \shownote      #1{#1}          \fi
\ifx \showarticletitle \undefined \def \showarticletitle #1{#1}   \fi
\ifx \showURL      \undefined \def \showURL       {\relax}        \fi
\providecommand\bibfield[2]{#2}
\providecommand\bibinfo[2]{#2}
\providecommand\natexlab[1]{#1}
\providecommand\showeprint[2][]{arXiv:#2}

\bibitem[Achiam et~al\mbox{.}(2023)]%
        {achiam2023gpt}
\bibfield{author}{\bibinfo{person}{Josh Achiam}, \bibinfo{person}{Steven Adler}, \bibinfo{person}{Sandhini Agarwal}, \bibinfo{person}{Lama Ahmad}, \bibinfo{person}{Ilge Akkaya}, \bibinfo{person}{Florencia~Leoni Aleman}, \bibinfo{person}{Diogo Almeida}, \bibinfo{person}{Janko Altenschmidt}, \bibinfo{person}{Sam Altman}, \bibinfo{person}{Shyamal Anadkat}, {et~al\mbox{.}}} \bibinfo{year}{2023}\natexlab{}.
\newblock \showarticletitle{Gpt-4 technical report}.
\newblock \bibinfo{journal}{\emph{arXiv preprint arXiv:2303.08774}} (\bibinfo{year}{2023}).
\newblock


\bibitem[Brown et~al\mbox{.}(2020)]%
        {brown2020language}
\bibfield{author}{\bibinfo{person}{Tom Brown}, \bibinfo{person}{Benjamin Mann}, \bibinfo{person}{Nick Ryder}, \bibinfo{person}{Melanie Subbiah}, \bibinfo{person}{Jared~D Kaplan}, \bibinfo{person}{Prafulla Dhariwal}, \bibinfo{person}{Arvind Neelakantan}, \bibinfo{person}{Pranav Shyam}, \bibinfo{person}{Girish Sastry}, \bibinfo{person}{Amanda Askell}, {et~al\mbox{.}}} \bibinfo{year}{2020}\natexlab{}.
\newblock \showarticletitle{Language models are few-shot learners}.
\newblock \bibinfo{journal}{\emph{Advances in neural information processing systems}}  \bibinfo{volume}{33} (\bibinfo{year}{2020}), \bibinfo{pages}{1877--1901}.
\newblock


\bibitem[Dong et~al\mbox{.}(2024)]%
        {dong-etal-2024-llm}
\bibfield{author}{\bibinfo{person}{Yijiang~River Dong}, \bibinfo{person}{Tiancheng Hu}, {and} \bibinfo{person}{Nigel Collier}.} \bibinfo{year}{2024}\natexlab{}.
\newblock \showarticletitle{Can {LLM} be a Personalized Judge?}. In \bibinfo{booktitle}{\emph{Findings of the Association for Computational Linguistics: EMNLP 2024}}. \bibinfo{publisher}{Association for Computational Linguistics}, \bibinfo{address}{Miami, Florida, USA}, \bibinfo{pages}{10126--10141}.
\newblock
\href{https://doi.org/10.18653/v1/2024.findings-emnlp.592}{doi:\nolinkurl{10.18653/v1/2024.findings-emnlp.592}}


\bibitem[Fu et~al\mbox{.}(2024)]%
        {fu-etal-2024-gptscore}
\bibfield{author}{\bibinfo{person}{Jinlan Fu}, \bibinfo{person}{See-Kiong Ng}, \bibinfo{person}{Zhengbao Jiang}, {and} \bibinfo{person}{Pengfei Liu}.} \bibinfo{year}{2024}\natexlab{}.
\newblock \showarticletitle{{GPTS}core: Evaluate as You Desire}. In \bibinfo{booktitle}{\emph{Proceedings of the 2024 Conference of the North American Chapter of the Association for Computational Linguistics: HLT (Long Papers)}}. \bibinfo{publisher}{Association for Computational Linguistics}, \bibinfo{address}{Mexico City, Mexico}, \bibinfo{pages}{6556--6576}.
\newblock
\href{https://doi.org/10.18653/v1/2024.naacl-long.365}{doi:\nolinkurl{10.18653/v1/2024.naacl-long.365}}


\bibitem[Gallegos et~al\mbox{.}(2024)]%
        {gallegos2024bias}
\bibfield{author}{\bibinfo{person}{Isabel~O Gallegos}, \bibinfo{person}{Ryan~A Rossi}, \bibinfo{person}{Joe Barrow}, \bibinfo{person}{Md~Mehrab Tanjim}, \bibinfo{person}{Sungchul Kim}, \bibinfo{person}{Franck Dernoncourt}, \bibinfo{person}{Tong Yu}, \bibinfo{person}{Ruiyi Zhang}, {and} \bibinfo{person}{Nesreen~K Ahmed}.} \bibinfo{year}{2024}\natexlab{}.
\newblock \showarticletitle{Bias and fairness in large language models: A survey}.
\newblock \bibinfo{journal}{\emph{Computational Linguistics}} \bibinfo{volume}{50}, \bibinfo{number}{3} (\bibinfo{year}{2024}), \bibinfo{pages}{1097--1179}.
\newblock


\bibitem[Gu et~al\mbox{.}(2025)]%
        {gu2025surveyllmasajudge}
\bibfield{author}{\bibinfo{person}{Jiawei Gu}, \bibinfo{person}{Xuhui Jiang}, \bibinfo{person}{Zhichao Shi}, \bibinfo{person}{Hexiang Tan}, \bibinfo{person}{Xuehao Zhai}, \bibinfo{person}{Chengjin Xu}, \bibinfo{person}{Wei Li}, \bibinfo{person}{Yinghan Shen}, \bibinfo{person}{Shengjie Ma}, \bibinfo{person}{Honghao Liu}, \bibinfo{person}{Saizhuo Wang}, \bibinfo{person}{Kun Zhang}, \bibinfo{person}{Yuanzhuo Wang}, \bibinfo{person}{Wen Gao}, \bibinfo{person}{Lionel Ni}, {and} \bibinfo{person}{Jian Guo}.} \bibinfo{year}{2025}\natexlab{}.
\newblock \showarticletitle{A Survey on LLM-as-a-Judge}.
\newblock \bibinfo{journal}{\emph{arXiv preprint arXiv:2411.15594}} (\bibinfo{year}{2025}).
\newblock
\urldef\tempurl%
\url{https://arxiv.org/abs/2411.15594}
\showURL{%
\tempurl}


\bibitem[Huang et~al\mbox{.}(2024)]%
        {huang2024foundation}
\bibfield{author}{\bibinfo{person}{Chengkai Huang}, \bibinfo{person}{Tong Yu}, \bibinfo{person}{Kaige Xie}, \bibinfo{person}{Shuai Zhang}, \bibinfo{person}{Lina Yao}, {and} \bibinfo{person}{Julian McAuley}.} \bibinfo{year}{2024}\natexlab{}.
\newblock \showarticletitle{Foundation models for recommender systems: A survey and new perspectives}.
\newblock \bibinfo{journal}{\emph{arXiv preprint arXiv:2402.11143}} (\bibinfo{year}{2024}).
\newblock


\bibitem[Jones et~al\mbox{.}(2021)]%
        {jones2021podcast}
\bibfield{author}{\bibinfo{person}{Rosie Jones}, \bibinfo{person}{Hamed Zamani}, \bibinfo{person}{Markus Schedl}, \bibinfo{person}{Ching-Wei Chen}, \bibinfo{person}{Sravana Reddy}, \bibinfo{person}{Ann Clifton}, \bibinfo{person}{Jussi Karlgren}, \bibinfo{person}{Helia Hashemi}, \bibinfo{person}{Aasish Pappu}, \bibinfo{person}{Zahra Nazari}, \bibinfo{person}{Longqi Yang}, \bibinfo{person}{Oguz Semerci}, \bibinfo{person}{Hugues Bouchard}, {and} \bibinfo{person}{Ben Carterette}.} \bibinfo{year}{2021}\natexlab{}.
\newblock \showarticletitle{Current Challenges and Future Directions in Podcast Information Access}. In \bibinfo{booktitle}{\emph{Proceedings of the 44th International ACM SIGIR Conference on Research and Development in Information Retrieval}}. \bibinfo{publisher}{ACM}, \bibinfo{address}{Virtual Event, Canada}, \bibinfo{pages}{1554--1565}.
\newblock
\urldef\tempurl%
\url{https://dblp.org/rec/conf/sigir/JonesZSC+21}
\showURL{%
\tempurl}


\bibitem[Lin et~al\mbox{.}(2025)]%
        {lin2025can}
\bibfield{author}{\bibinfo{person}{Jianghao Lin}, \bibinfo{person}{Xinyi Dai}, \bibinfo{person}{Yunjia Xi}, \bibinfo{person}{Weiwen Liu}, \bibinfo{person}{Bo Chen}, \bibinfo{person}{Hao Zhang}, \bibinfo{person}{Yong Liu}, \bibinfo{person}{Chuhan Wu}, \bibinfo{person}{Xiangyang Li}, \bibinfo{person}{Chenxu Zhu}, {et~al\mbox{.}}} \bibinfo{year}{2025}\natexlab{}.
\newblock \showarticletitle{How can recommender systems benefit from large language models: A survey}.
\newblock \bibinfo{journal}{\emph{ACM Transactions on Information Systems}} \bibinfo{volume}{43}, \bibinfo{number}{2} (\bibinfo{year}{2025}), \bibinfo{pages}{1--47}.
\newblock


\bibitem[Liu et~al\mbox{.}(2023)]%
        {liu-etal-2023-g}
\bibfield{author}{\bibinfo{person}{Yang Liu}, \bibinfo{person}{Dan Iter}, \bibinfo{person}{Yichong Xu}, \bibinfo{person}{Shuohang Wang}, \bibinfo{person}{Ruochen Xu}, {and} \bibinfo{person}{Chenguang Zhu}.} \bibinfo{year}{2023}\natexlab{}.
\newblock \showarticletitle{{G}-Eval: {NLG} Evaluation using Gpt-4 with Better Human Alignment}. In \bibinfo{booktitle}{\emph{Proceedings of the 2023 Conference on Empirical Methods in Natural Language Processing}}, \bibfield{editor}{\bibinfo{person}{Houda Bouamor}, \bibinfo{person}{Juan Pino}, {and} \bibinfo{person}{Kalika Bali}} (Eds.). \bibinfo{publisher}{Association for Computational Linguistics}, \bibinfo{address}{Singapore}, \bibinfo{pages}{2511--2522}.
\newblock
\href{https://doi.org/10.18653/v1/2023.emnlp-main.153}{doi:\nolinkurl{10.18653/v1/2023.emnlp-main.153}}


\bibitem[Sahoo et~al\mbox{.}(2025)]%
        {sahoo2025quantitativellmjudges}
\bibfield{author}{\bibinfo{person}{Aishwarya Sahoo}, \bibinfo{person}{Jeevana~Kruthi Karnuthala}, \bibinfo{person}{Tushar~Parmanand Budhwani}, \bibinfo{person}{Pranchal Agarwal}, \bibinfo{person}{Sankaran Vaidyanathan}, \bibinfo{person}{Alexa Siu}, \bibinfo{person}{Franck Dernoncourt}, \bibinfo{person}{Jennifer Healey}, \bibinfo{person}{Nedim Lipka}, \bibinfo{person}{Ryan Rossi}, \bibinfo{person}{Uttaran Bhattacharya}, {and} \bibinfo{person}{Branislav Kveton}.} \bibinfo{year}{2025}\natexlab{}.
\newblock \showarticletitle{Quantitative {LLM} Judges}. In \bibinfo{booktitle}{\emph{Proceedings of the 42nd International Conference on Machine Learning (ICML 2025)}}.
\newblock
\urldef\tempurl%
\url{https://arxiv.org/abs/2506.02945}
\showURL{%
\tempurl}
\newblock
\shownote{Spotlight, to appear}.


\bibitem[Su et~al\mbox{.}(2025)]%
        {su2025regen}
\bibfield{author}{\bibinfo{person}{Kun Su}, \bibinfo{person}{Krishna Sayana}, \bibinfo{person}{Hubert Pham}, \bibinfo{person}{James Pine}, \bibinfo{person}{Yuri Vasilevski}, \bibinfo{person}{Raghavendra Vasudeva}, \bibinfo{person}{Marialena Kyriakidi}, \bibinfo{person}{Liam Hebert}, \bibinfo{person}{Ambarish Jash}, \bibinfo{person}{Anushya Subbiah}, {et~al\mbox{.}}} \bibinfo{year}{2025}\natexlab{}.
\newblock \showarticletitle{REGEN: A Dataset and Benchmarks with Natural Language Critiques and Narratives}.
\newblock \bibinfo{journal}{\emph{arXiv preprint arXiv:2503.11924}} (\bibinfo{year}{2025}).
\newblock


\bibitem[Thakur et~al\mbox{.}(2025)]%
        {thakur2025judging}
\bibfield{author}{\bibinfo{person}{Aman~Singh Thakur}, \bibinfo{person}{Kartik Choudhary}, \bibinfo{person}{Venkat~Srinik Ramayapally}, \bibinfo{person}{Sankaran Vaidyanathan}, {and} \bibinfo{person}{Dieuwke Hupkes}.} \bibinfo{year}{2025}\natexlab{}.
\newblock \showarticletitle{Judging the Judges: Evaluating Alignment and Vulnerabilities in {LLM}s-as-Judges}.
\newblock \bibinfo{journal}{\emph{arXiv preprint arXiv:2406.12624}} (\bibinfo{year}{2025}).
\newblock
\urldef\tempurl%
\url{https://arxiv.org/abs/2406.12624}
\showURL{%
\tempurl}


\bibitem[Thomas et~al\mbox{.}(2024)]%
        {thomas2024llmsearchprefs}
\bibfield{author}{\bibinfo{person}{Paul Thomas}, \bibinfo{person}{Seth Spielman}, \bibinfo{person}{Nick Craswell}, {and} \bibinfo{person}{Bhaskar Mitra}.} \bibinfo{year}{2024}\natexlab{}.
\newblock \showarticletitle{Large Language Models Can Accurately Predict Searcher Preferences}. In \bibinfo{booktitle}{\emph{Proceedings of the 47th International ACM SIGIR Conference on Research and Development in Information Retrieval}}. \bibinfo{publisher}{ACM}, \bibinfo{address}{Washington DC, USA}, \bibinfo{pages}{1930--1940}.
\newblock
\href{https://doi.org/10.1145/3626772.3657707}{doi:\nolinkurl{10.1145/3626772.3657707}}


\bibitem[Vaswani et~al\mbox{.}(2017)]%
        {vaswani2017attention}
\bibfield{author}{\bibinfo{person}{Ashish Vaswani}, \bibinfo{person}{Noam Shazeer}, \bibinfo{person}{Niki Parmar}, \bibinfo{person}{Jakob Uszkoreit}, \bibinfo{person}{Llion Jones}, \bibinfo{person}{Aidan~N Gomez}, \bibinfo{person}{{\L}ukasz Kaiser}, {and} \bibinfo{person}{Illia Polosukhin}.} \bibinfo{year}{2017}\natexlab{}.
\newblock \showarticletitle{Attention is all you need}.
\newblock \bibinfo{journal}{\emph{Advances in neural information processing systems}}  \bibinfo{volume}{30} (\bibinfo{year}{2017}).
\newblock


\bibitem[Wang et~al\mbox{.}(2025)]%
        {wang2025user}
\bibfield{author}{\bibinfo{person}{Jianling Wang}, \bibinfo{person}{Yifan Liu}, \bibinfo{person}{Yinghao Sun}, \bibinfo{person}{Xuejian Ma}, \bibinfo{person}{Yueqi Wang}, \bibinfo{person}{He Ma}, \bibinfo{person}{Zhengyang Su}, \bibinfo{person}{Minmin Chen}, \bibinfo{person}{Mingyan Gao}, \bibinfo{person}{Onkar Dalal}, {et~al\mbox{.}}} \bibinfo{year}{2025}\natexlab{}.
\newblock \showarticletitle{User Feedback Alignment for LLM-powered Exploration in Large-scale Recommendation Systems}.
\newblock \bibinfo{journal}{\emph{arXiv preprint arXiv:2504.05522}} (\bibinfo{year}{2025}).
\newblock


\bibitem[Wang et~al\mbox{.}(2024a)]%
        {wang2024large}
\bibfield{author}{\bibinfo{person}{Jianling Wang}, \bibinfo{person}{Haokai Lu}, \bibinfo{person}{James Caverlee}, \bibinfo{person}{Ed~H Chi}, {and} \bibinfo{person}{Minmin Chen}.} \bibinfo{year}{2024}\natexlab{a}.
\newblock \showarticletitle{Large language models as data augmenters for cold-start item recommendation}. In \bibinfo{booktitle}{\emph{Companion Proceedings of the ACM Web Conference 2024}}. \bibinfo{pages}{726--729}.
\newblock


\bibitem[Wang et~al\mbox{.}(2024b)]%
        {wang2024llms}
\bibfield{author}{\bibinfo{person}{Jianling Wang}, \bibinfo{person}{Haokai Lu}, \bibinfo{person}{Yifan Liu}, \bibinfo{person}{He Ma}, \bibinfo{person}{Yueqi Wang}, \bibinfo{person}{Yang Gu}, \bibinfo{person}{Shuzhou Zhang}, \bibinfo{person}{Ningren Han}, \bibinfo{person}{Shuchao Bi}, \bibinfo{person}{Lexi Baugher}, {et~al\mbox{.}}} \bibinfo{year}{2024}\natexlab{b}.
\newblock \showarticletitle{Llms for user interest exploration in large-scale recommendation systems}. In \bibinfo{booktitle}{\emph{Proceedings of the 18th ACM Conference on Recommender Systems}}. \bibinfo{pages}{872--877}.
\newblock


\bibitem[Wei et~al\mbox{.}(2022)]%
        {wei2022chain}
\bibfield{author}{\bibinfo{person}{Jason Wei}, \bibinfo{person}{Xuezhi Wang}, \bibinfo{person}{Dale Schuurmans}, \bibinfo{person}{Maarten Bosma}, \bibinfo{person}{Fei Xia}, \bibinfo{person}{Ed Chi}, \bibinfo{person}{Quoc~V Le}, \bibinfo{person}{Denny Zhou}, {et~al\mbox{.}}} \bibinfo{year}{2022}\natexlab{}.
\newblock \showarticletitle{Chain-of-thought prompting elicits reasoning in large language models}.
\newblock \bibinfo{journal}{\emph{Advances in neural information processing systems}}  \bibinfo{volume}{35} (\bibinfo{year}{2022}), \bibinfo{pages}{24824--24837}.
\newblock


\bibitem[Xu et~al\mbox{.}(2025)]%
        {xu2025contextualjudgebench}
\bibfield{author}{\bibinfo{person}{Austin Xu}, \bibinfo{person}{Srijan Bansal}, \bibinfo{person}{Yifei Ming}, \bibinfo{person}{Semih Yavuz}, {and} \bibinfo{person}{Shafiq Joty}.} \bibinfo{year}{2025}\natexlab{}.
\newblock \showarticletitle{Does Context Matter? {ContextualJudgeBench} for Evaluating {LLM}-based Judges in Contextual Settings}. In \bibinfo{booktitle}{\emph{Proceedings of the 63rd Annual Meeting of the Association for Computational Linguistics (ACL 2025)}}.
\newblock
\urldef\tempurl%
\url{https://arxiv.org/abs/2503.15620}
\showURL{%
\tempurl}
\newblock
\shownote{To appear}.


\bibitem[Ye et~al\mbox{.}(2025)]%
        {ye2025justice}
\bibfield{author}{\bibinfo{person}{Jiayi Ye}, \bibinfo{person}{Yanbo Wang}, \bibinfo{person}{Yue Huang}, \bibinfo{person}{Dongping Chen}, \bibinfo{person}{Qihui Zhang}, \bibinfo{person}{Nuno Moniz}, \bibinfo{person}{Tian Gao}, \bibinfo{person}{Werner Geyer}, \bibinfo{person}{Chao Huang}, \bibinfo{person}{Pin-Yu Chen}, \bibinfo{person}{Nitesh~V. Chawla}, {and} \bibinfo{person}{Xiangliang Zhang}.} \bibinfo{year}{2025}\natexlab{}.
\newblock \showarticletitle{Justice or Prejudice? Quantifying Biases in {LLM}-as-a-Judge}. In \bibinfo{booktitle}{\emph{Proceedings of the International Conference on Learning Representations (ICLR 2025)}}.
\newblock
\urldef\tempurl%
\url{https://openreview.net/forum?id=3GTtZFiajM}
\showURL{%
\tempurl}
\newblock
\shownote{Poster}.


\bibitem[Zhang et~al\mbox{.}(2025)]%
        {zhang2025cold}
\bibfield{author}{\bibinfo{person}{Weizhi Zhang}, \bibinfo{person}{Yuanchen Bei}, \bibinfo{person}{Liangwei Yang}, \bibinfo{person}{Henry~Peng Zou}, \bibinfo{person}{Peilin Zhou}, \bibinfo{person}{Aiwei Liu}, \bibinfo{person}{Yinghui Li}, \bibinfo{person}{Hao Chen}, \bibinfo{person}{Jianling Wang}, \bibinfo{person}{Yu Wang}, {et~al\mbox{.}}} \bibinfo{year}{2025}\natexlab{}.
\newblock \showarticletitle{Cold-Start Recommendation towards the Era of Large Language Models (LLMs): A Comprehensive Survey and Roadmap}.
\newblock \bibinfo{journal}{\emph{arXiv preprint arXiv:2501.01945}} (\bibinfo{year}{2025}).
\newblock


\bibitem[Zheng et~al\mbox{.}(2023)]%
        {zheng2023mtbench}
\bibfield{author}{\bibinfo{person}{Lianmin Zheng}, \bibinfo{person}{Wei-Lin Chiang}, \bibinfo{person}{Ying Sheng}, \bibinfo{person}{Siyuan Zhuang}, \bibinfo{person}{Zhanghao Wu}, \bibinfo{person}{Yonghao Zhuang}, \bibinfo{person}{Zi Lin}, \bibinfo{person}{Zhuohan Li}, \bibinfo{person}{Dacheng Li}, \bibinfo{person}{Eric~P. Xing}, \bibinfo{person}{Hao Zhang}, \bibinfo{person}{Joseph~E. Gonzalez}, {and} \bibinfo{person}{Ion Stoica}.} \bibinfo{year}{2023}\natexlab{}.
\newblock \showarticletitle{Judging {LLM}-as-a-Judge with {MT‐Bench} and {Chatbot Arena}}. In \bibinfo{booktitle}{\emph{Advances in Neural Information Processing Systems 36 (NeurIPS 2023), Datasets and Benchmarks Track}}.
\newblock
\urldef\tempurl%
\url{https://proceedings.neurips.cc/paper/91f18a1287b398d378ef22505bf41832}
\showURL{%
\tempurl}


\bibitem[Zhu et~al\mbox{.}({[n.\,d.]})]%
        {zhujudgelm}
\bibfield{author}{\bibinfo{person}{Lianghui Zhu}, \bibinfo{person}{Xinggang Wang}, {and} \bibinfo{person}{Xinlong Wang}.} \bibinfo{year}{[n.\,d.]}\natexlab{}.
\newblock \showarticletitle{JudgeLM: Fine-tuned Large Language Models are Scalable Judges}. In \bibinfo{booktitle}{\emph{The Thirteenth International Conference on Learning Representations}}.
\newblock


\end{thebibliography}

\end{document}